\documentclass[reprint,aps,rapid,superscriptaddress,showpacs,floats,amsmath,amssymb]{revtex4-1}

\usepackage{color}
\usepackage{graphicx}
\usepackage{dcolumn}
\usepackage{bm}
\usepackage{float}
\usepackage{mciteplus}

\begin{document}

\title{Saturation of resistivity and Kohler's rule in Ni-doped
La$_{1.85}$Sr$_{0.15}$CuO$_{4}$ cuprate}
\author{A. Malinowski}
\affiliation{Institute of Physics, Polish Academy of Sciences,
al. Lotnik\'{o}w 32/46, 02-668 Warsaw, Poland}
\author{V. L. Bezusyy}
\affiliation{Institute of Physics, Polish Academy of Sciences,
al. Lotnik\'{o}w 32/46, 02-668 Warsaw, Poland}
\affiliation{International Laboratory of High Magnetic Fields and Low Temperatures,
ul. Gajowicka 95, 53-421 Wroclaw, Poland}
\author{P. Nowicki}
\affiliation{Institute of Physics, Polish Academy of Sciences,
al. Lotnik\'{o}w 32/46, 02-668 Warsaw, Poland}
\date{\today}

\begin{abstract}
We present the results of electrical transport measurements of
La$_{1.85}$Sr$_{0.15}$Cu$_{1-y}$Ni$_{y}$O$_{4}$  thin single-crystal films at magnetic fields
up to 9 T. Adding Ni impurity with strong Coulomb scattering potential to slightly underdoped cuprate
makes the signs of resistivity saturation at $\rho_{sat}$ visible in the measurement temperature window up to 350 K.
Employing the parallel-resistor formalism reveals that $\rho_{sat}$
is consistent with classical Ioffe-Regel-Mott limit and changes with
carrier concentration $n$ as $\rho_{sat}\propto 1/\sqrt{n}$. Thermopower measurements
show that Ni tends to localize mobile carriers,
decreasing their effective concentration as $n\!\cong0.15\!-\!y$.
The classical unmodified Kohler's rule is fulfilled for magnetoresistance in the nonsuperconducting
part of the phase diagram when applied to the ideal branch in the parallel-resistor model.
\end{abstract}

\maketitle
Increasing evidence for well-defined quasiparticles in underdoped cuprates seems to corroborate
a view that they are normal metals, only with small Fermi surface. Fermi-Dirac statistic underlying
the quantum oscillations \cite{Sebastian2010}, single-parameter - quadratic in energy $\omega$ and temperature $T$ - scaling
in optical conductivity $\sigma(\omega,T)$ (Ref.~\cite{Mirzaei2013}), $T^{2}$ resistivity behavior
extending over substantial $T$-region in clean systems \cite{Barisic2013a} and
fulfillment of typical for conventional metals Kohler's rule in magnetotransport \cite{Chan2014} are
observations in favor of Fermi-liquid scenario.

On the other hand, in cuprates with significant disorder manifested by large residual resistivity $\rho_{res}$,
pure $T^{2}$ resistivity dependence has not been reported so far as for Bi$_{2}$Sr$_{2}$CuO$_{6+\delta}$ \cite{Barisic2013a,Eisaki2004,Ando96}
or observed only at relatively narrow doping- and T-region as in La$_{2-x}$Sr$_{x}$CuO$_{4}$ (LSCO) \cite{Ando2004a,Barisic2013a}.
The clear violation of Kohler's scaling in underdoped LSCO and YBa$_{2}$Cu$_{3}$O$_{7}$ \cite{Harris95,Kimura96}
(although not necessarily meaning breakdown of Fermi-liquid description \cite{McKenzie98})
has served almost as a hallmark of their peculiar normal-state properties for two decades.

In contrast to overdoped cuprates where large cylindrical Fermi surface yields a carrier density
$n$=$p$+$1$ ($p$ being doping level) \cite{Plate2005,Vignolle2008,Vignolle2011}, the total volume of Fermi surface in underdoped systems
is a small fraction of the first Brillouin zone and corresponds to $n$=$p$
through Luttinger's theorem \cite{Doiron2007,Barisic2013,Badoux2016,Badoux2016a}.
This small $n$ should be reflected in zero-field transport. In normal metals, resistivity $\rho$ saturates in the vicinity of Ioffe-Regel-Mott limit
$\rho_{IRM}$ where elastic mean free path $l_{min}$ becomes comparable to interatomic distance \cite{Mott90,Wiesmann77}.
In cuprates, however, signs of saturation are seen at $\rho_{sat}$ much larger than $\rho_{IRM}$ calculated
from the semiclassical Boltzmann theory \cite{Orenstein90,Takagi92,Wang96,Takenaka2003}. Moreover, $\rho$(1000 K) ($\sim\!\rho_{sat}$) in LSCO changes as 1/$x$,
while for $n\propto x$ (Ref.~\cite{Takagi89}) the theory predicts $\rho_{sat}\propto1/\sqrt{x}$.

The above can be explained by breakdown of the quasiparticle picture due to strong inelastic scattering at high T
manifested by disappearing of a Drude peak in $\sigma(\omega)$ (Refs.~\cite{Takenaka2002,Takenaka2003,Gunnarsson2003,Calandra2003}).
In systems where impurity scattering dominates the carrier relaxation (quasiparticle decay) rate $1/\tau$, the Drude peak
is centered at $\omega$=0 regardless of how strong scattering becomes \cite{Mirzaei2013,Takenaka2003}.
Electron-electron interactions make $\tau$ frequency-dependent but
for Fermi-liquid-like $\omega^{2}$ dependence $\sigma(\omega)$ still peaks at $\omega$=0 (Ref.~\cite{Mirzaei2013}).
Thus large impurity-induced $\rho_{res}$ may facilitate approach
to Ioffe-Regel-Mott limit in dc ($\omega$=0) LSCO transport at lower $T$ before the spectral weight
is transferred to higher-energy excitations at larger $T$. Ni impurity is a good candidate
because its strong Coulomb scattering potential in the CuO$_{2}$ planes allows to achieve
large $\rho$ at moderately high $T$ \cite{Hudson2001,Tanabe2011}.

In this paper we report transport and thermopower measurements on
slightly underdoped $x$=0.15 LSCO with added Ni impurity. The obtained $\rho_{sat}$ corresponds to the classical value
for small Fermi surface and changes as $1/\sqrt{n}$. The Fermi-liquid quasiparticle picture holds in the nonsuperconducting part of the phase diagram, as
revealed by $\rho\!\propto\!T^{2}$ dependence and classical Kohler's rule for magnetoresistance, both hidden under the large resistivity of the system.

The 4-point transport measurements were carried out on the $c$-axis aligned single-crystal films grown on
isostructural LaSrAlO$_{4}$ substrate by the laser ablation method from
the polycrystalline targets \cite{Malinowski2011,Remark2015}. The thermopower, which is not sensitive to the
grain boundaries and porosity effects in cuprates \cite{Carrington94}, was measured on the samples cut from the targets.
\begin{figure}
\includegraphics[width=0.48\textwidth, trim= 0 0 0 0]
{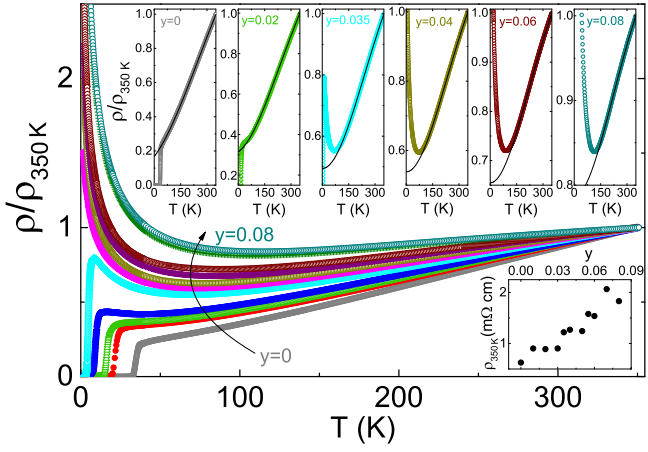}\caption{\label{Resistivity}(Color online) Temperature dependence of normalized resistivity
for a series of La$_{1.85}$Sr$_{0.15}$Cu$_{1-y}$Ni$_{y}$O$_{4}$ specimens. Their resistivities at $T$=350K are depicted in lower inset. Upper inset
shows the fits of Eq.~(\ref{ParRes}) to 150\! K-350\! K data for selected specimens.}
\end{figure}

\begin{figure}
\includegraphics[width=0.48\textwidth, trim= 0 0 0 0]
{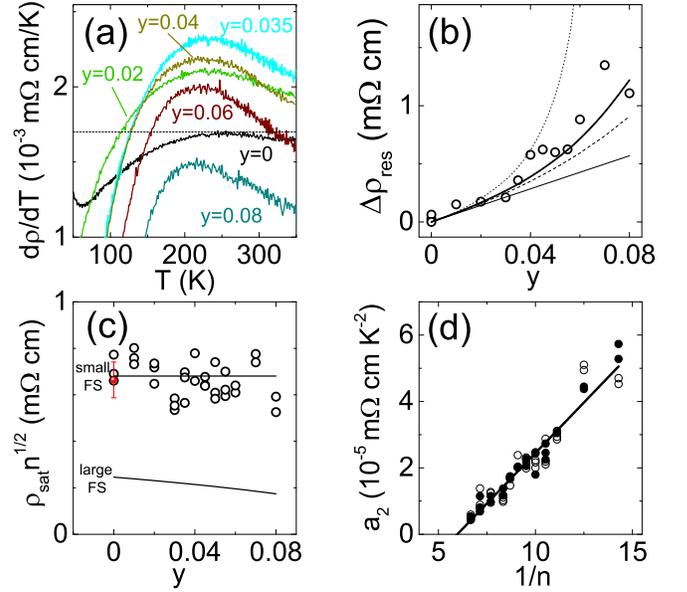}\caption{\label{FitResults}(Color online)
(a) Temperature derivatives of La$_{1.85}$Sr$_{0.15}$Cu$_{1-y}$Ni$_{y}$O$_{4}$ resistivity. The dashed horizontal line is a guide to the eye.
(b) The increase in residual resistivity of the samples with the smallest $\rho_{res}$ for a given $y$. The lines show
the unitarity limit assuming $n$=0.15$-\!y$ (thick line) and - for comparison - $n$=0.15 (thin line), $n$=0.15$-$0.7$y$ (dashed line) and hypothetical $n$=0.15$-$2$y$ (dotted line).
(c) The product $\rho_{sat}\sqrt n$ with the arithmetic mean $(\rho_{sat}\sqrt n)_{av}$ (red dot) for the 32 measured samples.
Solid lines show the expected $y$-dependence for small and large Fermi surface.
(d) The parameter $a_{2}$ as a function of inverse carrier concentration (open circles). Solid circles
mark the normalized values $a_{2}^{n}=a_{2}(\rho_{sat}\sqrt n)_{av}/(\rho_{sat}\sqrt{n})$ and solid line is the linear fit to them.}
\end{figure}

Figure \ref{Resistivity} shows the systematic change in $\rho(T)$  with Ni, from superconducting  $y$=0 specimen with midpoint $T_{C}$=34.6 K to $y$=0.08 one
exhibiting insulating behavior at low $T$.  In high $T$, a change of slope in portion of $\rho(T)$ curves that increases with $T$  foreruns
the approaching saturation. Variation of d$\rho$/d$T$, visible even in the $y$=0 data, authenticates the slope decreasing that becomes more pronounced with increasing $y$,
as can be seen in Fig.~\ref{FitResults}(a). Resistivity in this region is described extremely well by the parallel-resistor formula
\begin{equation}
\frac{1}{\rho(T)}=\frac{1}{\rho_{id}(T)}+\frac{1}{\rho_{sat}}=\frac{1}{a_{0}+a_{1}T+a_{2}T^{2}}+\frac{1}{\rho_{sat}}.\label{ParRes}
\end{equation}
The $\rho_{id}$ term is the ideal resistivity in the absence of saturation \cite{Wiesmann77}
and the additive-in-conductivity formalism stems from existence of the minimal scattering time $\tau_{min}$, equivalent to Ioffe-Regel-Mott limit,
which causes the shunt $\rho_{sat}$ to always influence $\rho$ in normal metals \cite{Gurvitch81,Hussey2003}.
The formula was used for overdoped LSCO \cite{Cooper2009} but with the large-Fermi-surface $\rho_{sat}$ value \cite{Hussey2003} as a fixed parameter.
The excellent fits of Eq.~(\ref{ParRes}) with all free parameters including $\rho_{sat}$ to $\rho(T)$ in 150 K-350 K interval are depicted in upper inset to Fig.~\ref{Resistivity}. Extending the fit interval downwards to lower $T$ for small $y$ does not change significantly the obtained fitting parameters, which are presented in Fig.~\ref{FitResults}(b-d).

The Ni-induced residual resistivity, calculated as $1/\rho_{res}$=$1/\rho_{sat}$+$1/a_{0}$,
accelerates with $y$ such that at large $y$ substantially exceeds s-wave scattering unitarity limit
$\Delta \rho_{res}$=$(\hbar/e^{2})(y/n)d$ for $n$=0.15=const, depicted as thin solid line in Fig.~\ref{FitResults}(b).
Here, $d$=$c/2\!\!\cong\!\!6.6$\!\! {\AA} is the average separation
of the CuO$_{2}$ planes in the La$_{1.85}$Sr$_{0.15}$Cu$_{1-y}$Ni$_{y}$O$_{4}$  films \cite{Remark2015}. To find the actual $n(y)$ dependence we carried out the thermopower measurements.

\begin{figure}
\includegraphics[width=0.48\textwidth, trim= 0 0 0 0]
{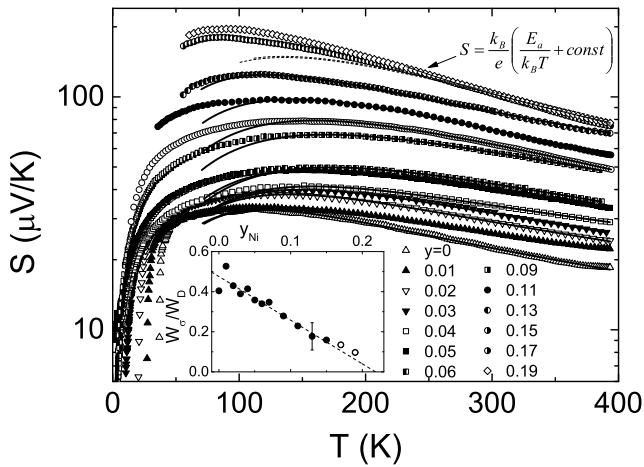}\caption{\label{SvsT}Temperature dependence of Seebeck coefficient for La$_{1.85}$Sr$_{0.15}$Cu$_{1-y}$Ni$_{y}$O$_{4}$  from $y$=0 (bottom) to $y$=0.19 (top).
The solid lines for $y\!\leq$0.15 and the dashed ones for $y\!>$0.15 are the best fits to the model from Ref.~\!\cite{Elizarova2000}.
The thin solid line is the best fit of thermally-activated transport formula for $y$=0.19.
Inset: The $W_{\sigma}/W_{D}$ ratio as a function of Ni doping. Dotted line is the best linear fit between $y$=0.02 and 0.15.}
\end{figure}

Ni doping increases the positive Seebeck coefficient $S$, as can be seen in Fig.~\ref{SvsT}. Taking into account the universal correlation
between $S$(290\! K) and hole concentration fulfilled in most of cuprate families \cite{Obertelli92,Honma2004,Kondo2005}, this strongly suggests decreasing
of carrier density with $y$. The $S(T)$ curves in La$_{1.85}$Sr$_{0.15}$Cu$_{1-y}$Ni$_{y}$O$_{4}$  retain the specific features of thermopower in underdoped cuprates:
the initial strong growth of $S$ with increasing $T$ is followed by a broad maximum and subsequent slight decrease in $S$ \cite{Remark2015d}.
We find that the phenomenological asymmetrical narrow-band model \cite{Elizarova2000} describes the experimental $S(T)$ curves
very well at high $T$, above their maximum at $T_{max}$.
In this model, a sharp density-of-states peak with the effective bandwidth $W_{D}$ is located near the Fermi level $E_{F}$ and the carriers
from the energy interval $W_{\sigma}$ are responsible for conduction. In addition, a shift $bW_{D}$ between the centers of $W_{D}$ and
$W_{\sigma}$ bands is assumed. The best fits of the formula determining $S$ in the model (Eq.~(1) in Ref.~\cite{Elizarova2000})
are shown as thick lines in Fig.~\ref{SvsT}. The discrepancies at low $T$ come from the limitations of the model derivated under the assumption $W_{D}\!\cong\!k_{B}T$.
Above $y$=0.15, the model also fails for larger $T$, well above $T_{max}$ ($W_{D}\!>\!380$ meV, while
$k_{B}T\!\cong\!26$ meV at 300 K). For $y$=0.17 and $y$=0.19,  $S$ changes as $\propto$1/$T$ above $\approx$250 K, consistent with the
formula $S(T)$=$(k_{B}/e)(E_{a}/(k_{B}T$+$const)$ indicative of polarons. The thermal activation energy $E_{a}$  estimated from the best fit
for $y$=0.19, $E_{a}$=32.4$\pm$0.2 meV, is in good agreement with Ref.~\cite{Zhiqiang98}.

At region of interest corresponding to high $T$, the asymmetrical narrow-band model works very well.
The obtained asymmetry factor $b\!=\!-$(0.06-0.08), although very small, is essential for good fits.
The ratio $F$ of $n$ to the total number od states $n_{DOS}$ is slightly above half-filling (0.51-0.53)
and thus consistent with the sign of $S$ and in good agrement with literature data for $x$=0.15 LSCO \cite{Elizarova2000,Zhou95}.
The Fermi level, $E_{F}$=$(F$--$0.5)W_{D}$--$bW_{D}$, crosses the conduction-band upper edge at $y\!\approx\!0.15$.
The band-filling $F$ does not show any obvious $y$-dependence. This means
that Ni does not change $n$ in the system (provided $n_{DOS}$ remains constant).
The primary effect of Ni doping appears to be \textit{localization} of existing mobile carriers, as revealed by
decreasing of $W_{\sigma}/W_{D}$ ratio with increasing $y$ (inset to Fig.~\ref{SvsT}). The ratio extrapolates to zero
at $y$=0.22$\pm$0.02, resulting in the average "localization rate" $\Delta n/\Delta y$=0.7$\pm$0.1 hole/Ni ion. Employing the simple two-band model with $T$-linear term \cite{Forro90},
where half-width of the resonance peak $\Gamma$ corresponds to the range of delocalized states,
results in the similar physical picture. We found that $\Gamma$ starts to decrease with increasing $y$ above 0.07 and approaches zero for $y$=0.17 (Ref.~\cite{Remark2015}).
The results are consistent with measurements of local distortion around Ni ions
suggesting trapping hole by \textit{each} Ni$^{2+}$  \cite{Hiraka2009} to create a well-localized Zhang-Rice doublet state \cite{Zhang88},
albeit indicate a slightly lower $\Delta n/\Delta y$.

Having established the effective mobile carrier concentration $n\!\cong\!0.15-y$, we can revert to Fig.~\ref{FitResults}b.
As indicated by thick solid line, scattering in the samples with the smallest $\rho_{res}$ is in the
unitarity limit for nonmagnetic impurity. Even assuming $\Delta n/\Delta y$=0.7 (and unitarity limit), at most only ~20\% of increase in $\rho_{res}$ for $y$=0.08
can be attributed to scattering on magnetic moments.
Thus, while our previous finding of spin-glass behavior in the system \cite{Malinowski2011}
undoubtedly indicates that the magnetic role of Ni ions in the spin-1/2 network of the CuO$_{2}$ planes
cannot be neglected, the scattering on Ni has a predominantly nonmagnetic origin \cite{Hudson2001}.

In Fig.~\ref{FitResults}c we show that the fitted  $\rho_{sat}$ agrees unexpectedly
with $\rho_{IRM}$ calculated from Boltzmann theory for the small Fermi surface with $n$ holes. Assuming a cylindrical surface with the height $2\pi/d$
and taking Ioffe-Regel-Mott condition as $l_{min}\!\approx\!a$ (i.e. $k_{F}l_{min}\!\approx\!\pi$, see Ref.~\cite{Hussey2004,Graham98}),
where $a$ is the lattice parameter in CuO$_{2}$ plane, one gets
$\rho_{IRM}^{small}$=$(\sqrt{2\pi}\hbar/e^{2})d/\sqrt{n}$=0.68$/\sqrt{n}$ m$\Omega$\! cm \cite{Gunnarsson2003}.
This is clearly distinguishable from the large Fermi surface case, $\rho_{IRM}^{large}$=0.68$/\sqrt{1+n}$, inapplicable to the system.
A simple formal statistics for all 32 measured samples shows that the product $\rho_{sat}\sqrt{n}$ for $n$=0.15$-y$ has
a distribution with mean\! $\cong$\! median and zero skewness \cite{Remark2015e}.

The precise location of the large-to-small Fermi surface transition on the phase diagram of cuprates
is still under debate. In the bismuth-based family, the expected linear relationship between $n$
estimated from $T_{C}$ and from Luttinger count is obtained only assuming large surface
from overdoped specimens down to $p\!\cong$0.145 inclusive \cite{He2014}.
In La$_{1.85}$Sr$_{0.15}$Cu$_{1-y}$Ni$_{y}$O$_{4}$, Ni doping effectively
moves the system towards smaller $p$ but the smooth evolution
of all the fitted $\rho(T)$ parameters down to $y$=0 points toward small Fermi surface at $p$=0.15.
This finding is consistent with recent Hall measurements in YBa$_{2}$Cu$_{3}$O$_{y}$
indicating that Fermi-surface reconstruction with decreasing doping ends sharply at $p$=0.16 (Ref.~\cite{Badoux2016}).

The $T$-linear coefficient in Eq.~(\ref{ParRes}) for three $y$=0 samples $\alpha_{1}$=0.93-1.0 $\mu\Omega$cm/K
is identical as that at $n_{cr}$=0.185$\pm$0.005 where a change in LSCO transport coefficients was found
when tracked from the overdoped side \cite{Cooper2009}.
Evidently, $\alpha_{1}$ is not sensitive to disappearing of the antinodal regions
during degradation/reconstruction of large Fermi surface into arcs/pockets.
The linear-in-T scattering is anisotropic in CuO$_{2}$ plane \cite{Abdel2006}
and its maximal level at ($\pi$,0) for $\alpha_{1}$=1 $\mu\Omega$cm/K is comparable \cite{Cooper2009} with Planckian dissipation limit \cite{Zaanen2004,Homes2004}.
De-coherence of quasiparticle states beyond this limit seems to be plausible explanation \cite{Cooper2009} of negligible
role of antinodal states in the conductivity.
While $\alpha_{1}$ vanishes for all specimens with $y\!\geqslant\!0.035$,
the $T^{2}$-coefficient $a_{2}$ changes linearly with 1/$n$ (compare Ref.~\cite{Barisic2013a}) in the whole studied $y$ range
(Fig.~\ref{FitResults}d). When extrapolated \textit{outside} accessible $n$,
$a_{2}$ approaches zero at $n_{cr}$=0.167$\pm$0.009. With such estimated error,
the result means that strictly linear $\rho(T)$ dependence (albeit masked by $\rho_{sat}$) in LSCO
is observed from the underdoped side only at optimum doping.
\textit{Interpreting} this as an indication of (antiferromagnetic) Quantum Critical Point remains speculative since
$a_{2}$ diverges at such a point \cite{Grigera01,Gegenwart02,Paglione03,Malinowski2005}.

\begin{figure*}
\includegraphics[width=1.00\textwidth, trim= 0 0 0 0]
{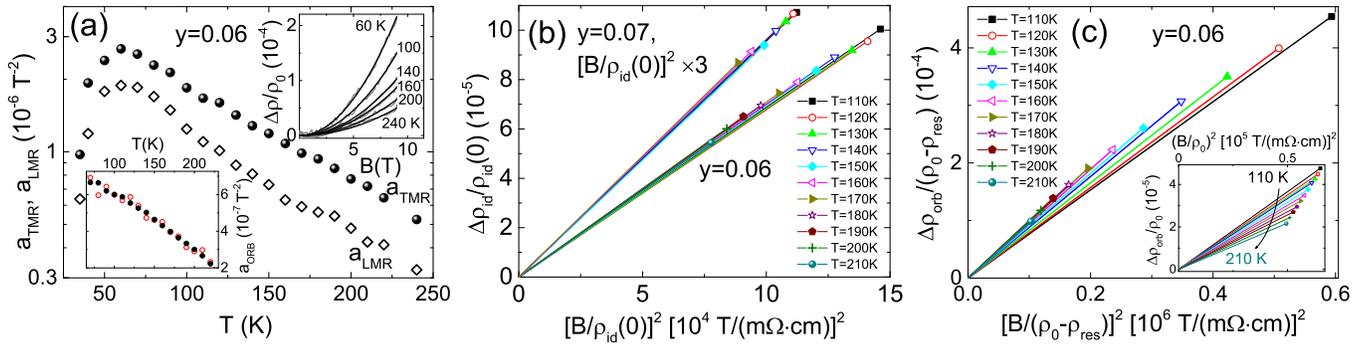}\caption{\label{Kohler}(Color online) Magnetoresistance of $y$=0.06 specimen. (a) The temperature dependence of the coefficients $a_{TMR}$ and $a_{LMR}$.
The orbital part $a_{ORB}$ (open symbols) and the result of its 5-point adjacent-averaging (solid symbols) are depicted in left inset.
Right inset shows $B^{2}$-fits to TMR data at selected $T$. (b) Kohler scaling for the ideal branch, together with that of
$y$=0.07 specimen for which abscissa scale is enlarged 3 times. (c) Scaling approach from Ref.~\cite{Chan2014} and
direct Kohler's-rule scaling attempt in inset.}
\end{figure*}
After using the parallel-resistor formalism in the zero-field transport description, we
will employ it to analysis of magnetoresistance in the sections below.
In the strongly overdoped cuprates magnetoresistance obeys Kohler's rule \cite{Kimura96}. The relative change of resistivity
in magnetic field B, $\Delta\rho/\rho_{0}$, is a unique function of $B/\rho_{0}$, where $\rho_{0}\!\equiv\!\rho(B$=0T).
In recent re-analysis of $x$=0.09 LSCO data from Ref.~\cite{Kimura96}, the modified Kohler's rule was proposed \cite{Chan2014}.
The isotemperature \textit{transverse} magnetoresistance vs. $B/\rho_{0}$ curves appeared to collapse onto single curve
when $\rho_{0}$ is replaced by $\rho_{0}\!-\!\rho_{res}$.
Here, we propose an alternative approach for considering the large LSCO resistivity in the magnetoresistance analysis.

At the lowest temperatures down to 2 K,
all non-superconducting samples exhibit large and negative in-plane ($I\!\!\parallel\!\!ab$) magnetoresistance, both in the transverse
(TMR, $B\!\!\perp\!\!ab$) and longitudinal (LMR, $B\!\!\parallel\!\!ab$) configuration [TMR($B$=9T)$\approx\!5\!\times\!10^{-2}$ at 2 K]
\cite{Remark2015}.
In the following, we focus on the high-$T$ region where, for the whole $y$ range studied, magnetoresistance in both configurations
is positive and two orders of magnitude smaller than in the low-$T$ region. The typical field dependence
of resistivity at various temperatures is displayed for $y$=0.06 specimen in the right inset to Fig.~\ref{Kohler}(a).
Above 35 K, $\rho$ increases as $B^{2}$. Below 60 K, the positive TMR begins to decrease with decreasing $T$
and smoothly evolves into negative one at low $T$. Similar behavior is observed for LMR, which constitutes
a significant portion of TMR (being equal to 60\% of TMR at $T$=150 K as an example) and thus may not be ignored
in the analysis. To obtain the orbital part, OMR, at $T\!\geqslant\!35$ K, we fitted
$\rho(B)$ with the form $[\Delta\rho(B)/\rho(0)]_{\text{TMR,LMR}}\!\!=\!\!a_{\text{TMR,LMR}}B^{2}$
and next calculated $a_{\text{ORB}}\!=\!a_{\text{TMR}}\!-\!a_{\text{LMR}}$.
The extracted coefficients are displayed in Fig.~\ref{Kohler}(a). A reliable and precise comparison of the various possible OMR scaling
requires moderate numerical smoothing without alternating the $a_{\text{ORB}}$ vs. $T$ dependence
[left inset to Fig.~\ref{Kohler}(a)]. Employing the smoothed  $a_{\text{ORB}}$ coefficients, $a_{\text{orb}}$,
OMR at any field $B$ can be calculated as OMR=$a_{\text{orb}}B^{2}$.

Clearly, Kohler's rule in La$_{1.85}$Sr$_{0.15}$Cu$_{1-y}$Ni$_{y}$O$_{4}$ is violated when applied directly to the measured OMR of the specimen
[inset to Fig.~\ref{Kohler}(c)]. Modification of the rule in the way described in Ref.~\cite{Chan2014} does not lead to any reasonable
scaling range \cite{Remark2015a}. The $\Delta\rho_{orb}/(\rho_{0}\!-\!\rho_{res})$ vs. $[B/(\rho_{0}\!-\!\rho_{res})]^{2}$ lines collapse one onto
another between 180 K and $T_{up}\!=\!200$ K, spanning only 10\% of $T_{up}$. Let's note that the existence of such a scaling - where the \textit{whole} absolute
resistivity change in field, $\Delta\rho_{orb}$, is related only to $T$-dependent $\rho_{el\text{-}ph}\!\equiv\!\rho_{0}\!-\!\rho_{res}$ part -
would mean in the classical picture that the field acts between the scattering events only on these carriers that scatter against phonons during the subsequent
scattering event and does not bend trajectories of those scattered against impurities.

The OMR analysis reveals that Kohler's rule in nonsuperconducting specimens is fulfilled in the ideal branch of the parallel-resistor model
where the influence of the shunt $\rho_{sat}$ is eliminated [Fig.~\ref{Kohler}(b)].
Interpreting the model in the spirit of the minimal $\tau_{0}$ leads to the assumption
that $\rho_{sat}$ is field-independent, at least in  the weak-field regime (where actually observed OMR$\propto\!\!B^{2}$
dependence is expected). Fitting of Eq.~(\ref{ParRes}) to $\rho(T)$ measured at various fields up to 9T
does not reveal any systematic change of $\rho_{sat}$ with $B$ \cite{Remark2015c}.
With $\rho_{sat}(B)\!\!=\!\!\rho_{sat}(0)$, OMR of the ideal branch,
$\Delta\rho_{id}/\rho_{id}(0)$
can be calculated from the measured quantities employing only one fitting parameter $\rho_{sat}(0)$.
The obtained $\Delta\rho_{id}/\rho_{id}(0)$ vs. $[B/(\rho_{id}(0)]^{2}$ curves from 110 K up to $T_{up}\!=\!210$ K  collapse to a single
temperature-independent line, spanning 50\% of $T_{up}$ \cite{Remark2015a}. The similar result was obtained
for $y$=0.07 specimen. Closer to the superconducting region of the phase diagram, for $y$=0.04, the scaling interval in $\rho_{id}(B,T)$ is much smaller (140K-160K)
but larger difference between $\rho_{sat}$ and $\rho_{res}\!\!\approx\!\!0.3\rho_{sat}$
emphasizes the difference between the possible OMR scalings and the scaling approach illustrated in Fig.~\ref{Kohler}(c) fails completely.

Concluding, the signs of resistivity  saturation both at the value $\rho_{IRM}$ and with
$n$-dependence from Boltzmann theory reflect metallic-like character of transport despite
small volume of Fermi surface and strong disorder in underdoped LSCO.
A fully quantum-mechanical explanation of saturation is still lacking
\cite{Calandra2002,Gunnarsson2003,Hussey2004,Allen2002,Werman2016}.
Evidently however, strong electron-electron interactions \cite{Remark2016}
do not invalidate the $\rho_{IRM}$ limit \cite{Calandra2003}.
The Ni-induced order-of-magnitude increase of $a_{0}/\rho_{sat}$ ratio
leaves the resulting $\rho_{IRM}^{small}\sqrt{n}$ intact.
The revealed omnipresence of $\tau_{min}$ even in bad metals points towards universality of
the Ioffe-Regel-Mott criterion \cite{Hussey2004}
rather than accidental-only agreement between $\rho_{IRM}$ and saturation \cite{Calandra2003}.

The known huge increase of $\rho$ in LSCO outside our $T$-measurement window means the breakdown of quasiparticle description.
At lower $T$, the transport in La$_{1.85}$Sr$_{0.15}$Cu$_{1-y}$Ni$_{y}$O$_{4}$
has a coherent description in the framework of Fermi-liquid theory, where Kohler's rule is derived under
single-relaxation-time approximation with the assumption of small $\tau$-anisotropy over Fermi surface \cite{Ziman72,Kontani2008}.
Fulfillment of the rule when $\rho_{0}$ is changed by changing temperature
indicates nearly $T$-independent frequency distribution of the phonons involved \cite{Pippard89}
and is consistent with Fermi-arc length in cuprates being constant \cite{Kaminski2013,Kaminski2014} rather than decreasing
with decreasing temperature \cite{Kanigel2006}.
Ni doping can restore antiferromagnetic fluctuations \cite{Tsutsui2009,Malinowski2011} that give additional $T$-dependence in nearly-antiferromagnetic-metals
magnetoresistance via correlation length $\xi_{AF}(T)$, OMR$\propto$$\xi_{AF}^{4}(T)\rho_{0}^{-2}$ (Ref.~\cite{Kontani2008}). Thus fulfillment of conventional
Kohler's rule suggests $T$-independent $\xi_{AF}$ within the framework of this theory.

In summary, the typical for normal metals parallel-resistor formalism,
employed for analysis of La$_{1.85}$Sr$_{0.15}$Cu$_{1-y}$Ni$_{y}$O$_{4}$ transport,
reveals resistivity saturation at the value expected from Boltzmann theory
for the small Fermi surface and fulfilment of unmodified Kohler's rule in the nonsuperconducting part
of the  phase diagram.

\begin{acknowledgments}
This research was partially performed in the NanoFun laboratory co-financed by the ERDF Project
NanoFun POIG.02.02.00-00-025/09.
\end{acknowledgments}


\begin{thebibliography}{69}
\bibitem[{Rem({\natexlab{c}})}]{Remark2015e}
\bibinfo{note}{The linear fit to $\rho_{sat}\sqrt{n}(y)$ for $n\!=\!0.15\!-\!y$
  has a small nonzero slope that vanishes for $n\!=\!0.15\!-\!0.7y$, but the
  result should be taken with some caution because of relatively large
  scattering of data.}\relax
\mciteBstWouldAddEndPunctfalse
\mciteSetBstMidEndSepPunct{\mcitedefaultmidpunct}
{}{\mcitedefaultseppunct}\relax
\EndOfBibitem
\bibitem[{Rem({\natexlab{d}})}]{Remark2015a}
\bibinfo{note}{The $T$-range of the scaling has been found semiquantitatively
  as a maximal one for which adding a next measured $T$-point outside this
  range doesn't broaden the existing angle between the outermost lines in the
  plot.}\relax
\mciteBstWouldAddEndPunctfalse
\mciteSetBstMidEndSepPunct{\mcitedefaultmidpunct}
{}{\mcitedefaultseppunct}\relax
\EndOfBibitem
\bibitem[{Rem({\natexlab{e}})}]{Remark2015c}
\bibinfo{note}{The larger field is needed to unmask the possible $B$-dependence
  of $\rho_{sat}$ (when treated as a formal fitting parameter) that may be
  obscured inside small magnetoresistance of the system}\relax
\mciteBstWouldAddEndPuncttrue
\mciteSetBstMidEndSepPunct{\mcitedefaultmidpunct}
{\mcitedefaultendpunct}{\mcitedefaultseppunct}\relax
\EndOfBibitem
\end{thebibliography}
\end{document}